\documentclass[12pt]{article}

  \def\fH{{\cal H}}

  \def\fX{{\cal X}}

 \def\fW{{\cal W}}
\def\Tr{\hbox{\rm Tr}\,}
\def\({\Big(}
\def\){\Big)}

\newtheorem{Lem}{Lemma}[section]
\newtheorem{Def}[Lem]{Definition}
\newtheorem{The}[Lem]{Theorem}
\newtheorem{Prop}[Lem]{Proposition}

\usepackage{amssymb}
\usepackage{latexsym}

\begin{document}

\title{Concavity of the auxiliary function appearing in quantum reliability function in classical-quantum channels}
\author{Jun Ichi Fujii \footnote{Department of Arts and Sciences (Information Science), Osaka Kyoiku University, Kashiwara-city, Osaka, 582-8582, Japan. e-mail: fujii@cc.osaka-kyoiku.ac.jp}, \, Ritsuo Nakamoto \footnote{Faculty of Engineering, Ibaraki University, Hitachi-city, 316-8511, Japan. e-mail: nakamoto@base.ibaraki.ac.jp}\ \, and Kenjiro Yanagi \footnote{Department of Applied Science, Faculty of Engineering, Yamaguchi University, Ube city, 755-8611, Japan. e-mail: yanagi@yamaguchi-u.ac.jp}} 
\date{}
\maketitle
{\bf Abstract.} Concavity of the auxiliary function which appears in the random coding exponent as the lower bound of the quantum reliability function for general quantum states is proven for $0\leq s \leq 1$. 

\vspace{3mm}
{\bf Running Head: } Concavity of the auxiliary function in quantum reliability function

\vspace{3mm}
\vspace{3mm}

\section{Introduction}

In quantum information theory, it is important to study the properties of the 
auxiliary function $E_q (\pi ,s)$, which will be defined in the below, 
appearing in the lower bound with respect to the random coding in the 
reliability function for general quantum states. In classical information 
theory \cite{Gal}, the random coding exponent $E^c_r(R)$, the lower bound of 
the reliability function, is defined by
$$
E^c_r(R) = \max_{p,s}\left[ E_c(p,s) -s R \right]. 
$$
As for the classical auxiliary function $E_c(p,s)$, it is well-known the 
following properties \cite{Gal}. 
\begin{itemize}
\item[(a)] $E_c(p,0) =0 .$
\item[(b)] $\displaystyle{\frac{\partial E_c(p,s)}{\partial s} \vert _{s=0} = I(X;Y),}$ where $I(X;Y)$ presents the classical mutual information.
\item[(c)] $E_c(p,s) > 0\,\,(0<s\leq 1)$. $E_c(p,s) < 0\,\,(-1< s \leq 0)$.
\item[(d)] $\displaystyle{\frac{\partial E_c(p,s)}{\partial s} > 0}$, $(-1 < s\leq 1)$.
\item[(e)] $\displaystyle{\frac{\partial^2 E_c(p,s)}{\partial s^2} \leq 0}$, $(-1 < s \leq 1)$.
\end{itemize}

In quantum case, the corresponding properties to (a),(b),(c) and (d) have been 
shown in \cite{ON,Hol2}.  Also the concavity of the auxiliary function 
$E_q (\pi ,s)$ is shown in the case when the signal states are pure 
\cite{Hol3}, and when the expurgation method is adopted \cite{Hol2}. 
However, for general signal states, the concavity of the function 
$E_q(\pi ,s)$ which corresponds to (e) in the above has remained as an open 
question \cite{ON} and still unsolved conjecture \cite{Hol2}. 


\section{Quantum reliability function}

The reliability function of classical-quantum channel is defined by
\begin{equation} \label{eq:num1}
E(R) \equiv - \liminf_{n\to \infty} \frac{1}{n} \log P_e(2^{nR},n),\quad 0<R<C,
\end{equation}
where $C$ is a classical-quantum capacity, $R$ is a transmission rate $R= \frac{\log_2 M}{n}$ ($n$ and $M$ represent the length and the number of the code 
words, respectively), $P_e(M,n)$ can be taken any minimal error probabilities of $\min_{\fW,\fX} \bar{P}(\fW,\fX)$ or $\min_{\fW,\fX} P_{\max}(\fW,\fX)$. These error probabilities are defined by 
\begin{eqnarray*}
\bar{P}(\fW,\fX) &=& \frac{1}{M}\sum_{j=1}^MP_j(\fW,\fX), \\
P_{\max}(\fW,\fX) &=& \max_{1\leq j \leq M}P_j(\fW,\fX), 
\end{eqnarray*}
where
$$
P_j(\fW,\fX) = 1- \Tr S_{w^j}X_j
$$
is the usual error probability associated with the positive operator valued 
measurement $\fX =\left\{X_j\right\}$ satisfying $\sum_{j=1}^M X_j \leq I$.  
Here we note $S_{w^j}$ represents the density operator corresponding to the 
code word $w^j$ choosen from the code(blook) $\fW =\left\{w^1,w^2,\cdots ,w^M\right\}$.  For details, see \cite{Hol1,ON,Hol2}. \\
The lower bound for the quantum reliability function defined in 
Eq.(\ref{eq:num1}), when we use random coding, is given by 
$$
E(R) \geq E^q_r(R) \equiv \max _{\pi} \sup _{0 <s \leq 1}\left[E_q \left(\pi ,s\right)-s R\right],
$$
where $\pi = \left\{ \pi_1,\pi_2,\cdots ,\pi_a \right\}$ is {\it a priori} 
probability distribution satisfying $\sum_{i=1}^a\pi_i=1$ and 
\begin{equation} \label{eq:num2}
E_q \left(\pi ,s\right) = -\log \Tr \left[ \left(\sum_{i=1}^a \pi_i S_i^{\frac{1}{1+s}} \right)^{1+s}\right], 
\end{equation}
where each $S_i$ is a non-degenerate density operator which corresponds to the 
output state of the classical-quantum channel $i \to S_i$ from the set of the 
input alphabet $A=\left\{1,2,\cdots ,a\right\}$ to the set of the output 
quantum states in the Hilbert space $\fH$. For the problem stated in previous section, a sufficient condition on concavity of the auxiliary function was given  in the following. 

\begin{Prop}[\cite{FYK}] 
If the trace inequality 
\begin{equation} \label{eq:num3}
\Tr\left[A(s)^s \left\{\sum_{j=1}^a 
\pi_jS_j^{\frac{1}{1+s}}\left(\log S_j^{\frac{1}{1+s}}\right)^2\right\}-A(s)^{-1+s}
\left\{\sum_{j=1}^a \pi_j H\left(S_j^{\frac{1}{1+s}} \right)
\right\}^2\right] \geq 0.
\end{equation}
holds for any real number $s \,\,(-1 < s \leq 1)$, any density matrices 
$S_i (i=1,\cdots ,a)$ and any probability distributions 
$\pi = \left\{\pi_i\right\}_{i=1}^a$, 
under the assumption that $A(s) \equiv \sum_{i=1}^a \pi_i S_i^{\frac{1}{1+s}}$ 
is invertible, then the auxiliary function $E_q  \left(\pi ,s\right)$ defined 
by Eq.(\ref{eq:num2}) is concave for all $s \, \, (-1 < s \leq 1)$. Where $H(x) = -x\log x$ is the matrix entropy. \vspace{3mm} \\ 
\end{Prop}

We note that our assumption \lq\lq $A(s)$ is invertible" is not so special 
condition, because $A(s)$ becomes invertible if we have one invertible $S_i$ 
at least. Moreover, we have the possibility such that $A(s)$ becomes 
invertible even if all $S_i$ is not invertible for all $\pi_i \neq 0$. 

In \cite{YFK}, Yanagi, Furuichi and Kuriyama proved the concavity of 
$E_q \left(\pi ,s\right)$ in the special case $a = 2$ with 
$\pi_1 = \pi_2 = \frac{1}{2}$ under the assumption that the dimension of 
$\fH$ is two by proving the trace inequality (\ref{eq:num3}). And recently 
in \cite{Fuj}, Fujii proved (\ref{eq:num3}) in the case 
$a = 2$ with $\pi_1 = \pi_2 = \frac{1}{2}$ under any dimension of $\fH$. 
In this paper we prove (\ref{eq:num3}) for any $a$ under any dimension of 
$\fH$. Then it is shown that $E_q(\pi,\cdot)$ is concave on $[0,1]$.


\section{Main Results}

We need several results in order to state the main theorem. 

\begin{Def}[\cite{Bou1},\cite{Bou2}] 
Let $f, g$ be real valued continuous functions.  Then $(f,g)$ is called 
a monotone (resp. antimonotone) pair of functions on the domain 
$D \subset \mathbb{R}$ if 
$$
(f(a)-f(b))(g(a)-g(b)) \geq 0 \; \; (resp. \leq)
$$
for any $a, b \in D$.
\end{Def}

\begin{Prop}[\cite{Bou1},\cite{Bou2},\cite{Fuj}]
If $(f,g)$ is a monotone (resp. antimonotone) pair, then 
$$
\Tr[f(A)Xg(A)X] \leq \Tr[f(A)g(A)X^2] \; \; (resp. \geq)
$$
for selfadjoint matrices $A$ and $X$ whose spectra are included in $D$.
\label{prop:proposition3.2}
\end{Prop}

\begin{Prop}[\cite{Fuj}] 
Let $S_1^{\frac{1}{1+s}} = A, S_2^{\frac{1}{1+s}} = B$ and 
$\pi_1 = \pi_2 = \frac{1}{2}$. Then 
$$
\Tr[(A+B)^s(A(\log A)^2+B(\log B)^2)-(A+B)^{s-1}(A\log A+B\log B)^2] \geq 0,
$$
for $s \geq 0$.
\end{Prop}

Now we state the main theorem. 

\begin{The}
Let $S_i^{\frac{1}{1+s}} = A_i \, (i = 1,\ldots,a)$. Then 
$$
\Tr \left[\(\sum_{k=1}^a \pi_k A_k\)^s \sum_{i=1}^a \pi_iA_i(\log A_i)^2-\(\sum_{k=1}^a \pi_k A_k\)^{s-1}\(\sum_{i=1}^a \pi_i A_i \log A_i\)^2 \right] \geq 0, 
$$
for $s \geq 0$.
\end{The}

\begin{flushleft}
{\bf Proof}.  We recall the following Jensen's inequality 
(e.g. \cite{HP,FF}): If $\displaystyle\sum_{i=1}^a C_i^*C_i = I$, then 
\end{flushleft}
$$
\sum_{i=1}^a C_i^*X_i^2C_i \geq \(\sum_{i=1}^a C_i^*X_iC_i\)^2
$$
holds for any Hermitian operators $X_i$, since $f(x) = x^2$ is operator 
convex on any interval. We put 
$$
X_i = \log A_i, \; \; C_i = (\pi_iA_i)^{1/2}\(\sum_{k=1}^a \pi_kA_k\)^{-1/2} 
\; (i = 1,2,\ldots,a).
$$
Since $\displaystyle{\sum_{i=1}^a C_i^*C_i = I}$, we have 
\begin{eqnarray*}
&   & \sum_{i=1}^a\(\sum_{k=1}^a \pi_kA_k\)^{-1/2}(\pi_iA_i)^{1/2}(\log A_i)^2(\pi_iA_i)^{1/2}\(\sum_{k=1}^a \pi_kA_k\)^{-1/2} \\
& \geq & \left(\sum_{i=1}^a\(\sum_{k=1}^a \pi_kA_k\)^{-1/2}(\pi_iA_i)^{1/2}\log A_i (\pi_i A_i)^{1/2}\(\sum_{k=1}^a \pi_kA_k\)^{-1/2} \right)^2.
\end{eqnarray*}
And so we have 
\begin{eqnarray*}
&   & (\sum_{k=1}^a \pi_kA_k)^{-1/2}\sum_{i=1}^a (\pi_iA_i)^{1/2}(\log A_i)^2(\pi_iA_i)^{1/2}(\sum_{k=1}^a \pi_kA_k)^{-1/2} \\
& \geq & \left(\(\sum_{k=1}^a \pi_kA_k\)^{-1/2}\(\sum_{i=1}^a \pi_iA_i \log A_i\)\(\sum_{k=1}^a \pi_kA_k\)^{-1/2} \right)^2.
\end{eqnarray*}
Hence it follows that 
\begin{eqnarray*}
&   & \sum_{i=1}^a (\pi_iA_i)^{1/2}(\log A_i)^2(\pi_iA_i)^{1/2} \\
& \geq & \(\sum_{i=1}^a \pi_iA_i \log A_i\)\(\sum_{k=1}^a \pi_kA_k\)^{-1}\(\sum_{i=1}^a \pi_iA_i \log A_i\).
\end{eqnarray*}
Then we have 
\begin{eqnarray*}
&   & \(\sum_{k=1}^a \pi_kA_k\)^{s/2}\sum_{i=1}^a \pi_i A_i(\log A_i)^2\(\sum_{k=1}^a \pi_kA_k\)^{s/2} \\
& \geq & \(\sum_{k=1}^a \pi_kA_k\)^{s/2}\(\sum_{i=1}^a \pi_iA_i \log A_i\)\(\sum_{k=1}^a \pi_kA_k\)^{-1}\(\sum_{i=1}^a \pi_iA_i \log A_i\)\(\sum_{k=1}^a \pi_kA_k\)^{s/2}.
\end{eqnarray*}
Thus 
\begin{eqnarray*}
&   & \Tr \left[\(\sum_{k=1}^a \pi_kA_k\)^s \sum_{i=1}^a \pi_iA_i (\log A_i)^2 \right] \\
& \geq & \Tr \left[\(\sum_{k=1}^a \pi_kA_k\)^s\(\sum_{i=1}^a \pi_iA_i \log A_i\)\(\sum_{k=1}^a \pi_kA_k\)^{-1}\(\sum_{i=1}^a \pi_iA_i \log A_i\) \right].
\end{eqnarray*}
Since $f(x) = x^s \; (s \geq 0)$ and $g(x) = x^{-1}$, it is clear that 
$(f,g)$ is antimonotone  pair. By Proposition \ref{prop:proposition3.2}, 
$$
\Tr \left[\(\sum_{k=1}^a \pi_k A_k\)^s \sum_{i=1}^a \pi_iA_i(\log A_i)^2-\(\sum_{k=1}^a \pi_k A_k\)^{s-1}\(\sum_{i=1}^a \pi_i A_i \log A_i\)^2 \right] \geq 0. 
$$
\ \hfill q.e.d.

\vspace{1cm}
We conclude that in this paper we finally solved the open problem given by 
\cite{Hol2} \cite{ON} that $E_q(\pi,\cdot)$ is concave on $[0,1]$.

\end{document}